\documentstyle[12pt]{article}
\newcommand{\INST}[1]{{\protect\thanks{#1}}\ }
\author{}
\title{ {\small{\rightline {\hfill LBL--33031} \vspace{0.8cm}}}
Relativistic Mean--Field Calculations of $\Lambda $ and $\Sigma $ Hypernuclei
\INST {
This work was supported by the Director, Office of Energy Research, Office of
High Energy and Nuclear Physics, Division of Nuclear Physics, of the U.S.
Department of Energy under Contract No. DE--AC03--76SF00098. }}
\renewcommand{\theequation}{\arabic{equation}}
\begin{document}

\pagenumbering{roman}

\maketitle
{\centerline {\bf N.K.~Glendenning}}
\medskip
{\centerline {\it Nuclear Science Division, Lawrence Berkeley Laboratory,}}
{\centerline {\it University of California, Berkeley, California 94720,}}
{\centerline {\it U.S.A.}}
\bigskip
{\centerline {\bf D.~Von-Eiff, M.~Haft, H.~Lenske, and M.K.~Weigel}}
\medskip
{\centerline {\it Sektion Physik der Ludwig--Maximilians--Universit\"at
                  M\"unchen,}}
{\centerline {\it Theresienstrasse 37/III, W--8000 M\"unchen 2,}}
{\centerline {\it Federal Republic of Germany}}

\newpage

{{\centerline{\large
Relativistic Mean--Field Calculations of $\Lambda $ and $\Sigma $ Hypernuclei
}}~\\[3ex]
{\centerline {\bf N.K.~Glendenning}}
{\centerline {\bf D.~Von-Eiff, M.~Haft, H.~Lenske, and M.K.~Weigel}}~\\[2ex]

\begin{abstract}
Single--particle spectra of $\Lambda $ and $\Sigma $ hypernuclei are calculated
within a relativistic mean--field theory. The  hyperon couplings
used are
compatible with the $\Lambda $ binding in saturated nuclear matter,
neutron-star masses and experimental data on $\Lambda $ levels in hypernuclei.
Special attention is devoted to the spin-orbit potential for the hyperons and
the influence of the $\rho $-meson field (isospin dependent interaction).
\end{abstract}
\centerline{{\bf PACS} number: 21.80.+a}

\newpage

\pagenumbering{arabic}

{{\centerline{\large
Relativistic Mean--Field Calculations of $\Lambda $ and $\Sigma $ Hypernuclei
}}~\\[3ex]
{\centerline {\bf N.K.~Glendenning}}
{\centerline {\bf D.~Von-Eiff, M.~Haft, H.~Lenske, and M.K.~Weigel}}~\\[2ex]

Nowadays the hypernuclear physics are of great interest in many branches of
physics. Of particular importance is the understanding of strange particles
in baryonic matter, since many questions in heavy-ion physics and astrophysics
are related to the effect of strangeness in matter. Experimentally and
theoretically such problems have been mainly studied for the $\Lambda $
hyperon, because it has zero isospin and charge. For $\Lambda $
particles theoretical approaches
for the spectroscopy range from nonrelativistic models \cite{one} to the
relativistic Hartree approximation (RHA) \cite{two,three}. For the RHA one uses
a Lagrangian with effective $\Lambda $ couplings to the $\sigma $-
and $\omega $-meson fields.
In Ref. \cite{three} the $\Lambda $ coupling constants (i.e. their relative
strength to the corresponding nucleon couplings $x_\sigma =g_{\Lambda \sigma }
/g_\sigma $ and $x_\omega =g_{\Lambda \omega }/g_\omega $) have been fitted to
the experimental $\Lambda $ hypernuclei spectra. However, treating $x_\sigma $
and $x_\omega $ as independent parameters leads to a highly uncertain
determination (correlation errors up to $\pm $65\% in Ref. \cite{three}).

On the other hand, the contribution of the hyperons strongly influences
the mass of neutron-stars. In a recent publication \cite{four} Glendenning
and Moszkowski related the scalar and vector couplings of the $\Lambda $
hyperon to its empirical binding in saturated nuclear matter \cite{one}
and thereby obtained compatibility of this binding energy with maximum
neutron-star masses. In fact, the large correlation found in the least
squares fit mentioned above \cite{three} reflects this relation of
$x_\sigma $ and $x_\omega $ to the $\Lambda $ binding in nuclear
matter. In summary, one finds that (1) neutron-star masses, (2) the
$\Lambda $ binding in saturated nuclear matter, and (3) $\Lambda $ levels
in hypernuclei are mutually compatible and rather narrowly constraining the
$\Lambda $ couplings.

Concerning $\Sigma $ hypernuclei, up to now the experimental situation
is not satisfactory. $\Sigma $ hypernuclear production has been investigated
at CERN, and later at Brookhaven and KEK, but the statistical accuracy
of the available data is not very good, because of the strong conversion
the $\Sigma $ undergoes in the nucleus ($\Sigma N \rightarrow \Lambda N$).
The controversial evidence for narrow $\Sigma $ states ($\Gamma < 5-10$~MeV)
is reviewed in Ref. \cite{five}. Therefore, in theoretical investigations
including $\Sigma $ hyperons
one usually assumes an universal hyperon coupling; i.e. all hyperons in the
lowest octet have the same coupling as the $\Lambda $ \cite{four,six}.
The prospects for significant advances in high resolution hypernuclear
spectroscopy at CEBAF or at future facilities such as the proposed
PILAC and KAON are discussed in Ref. \cite{seven}.

It is the aim of this contribution to analyze $\Lambda $ hypernuclei under
consideration
of the constraints (1)--(3) mentioned above and to extend such an investigation
to $\Sigma $ hypernuclei.

For the nucleonic sector we use the common nuclear field theory Lagrangian
including the
nucleon couplings to the $\sigma $-, $\omega $-, and $\rho $-meson fields
\cite{eight} plus phenomenological  $\sigma $-selfinteractions \cite{nine}.
For the three charge states of the $\Sigma $ hyperon we write the following
Lagrangian \cite{ten,eleven}:
\begin{eqnarray}
{\cal L}& = & \sum _\Sigma \bar \psi _\Sigma \left( i\gamma _\mu \partial
^\mu -M_\Sigma +g_{\Sigma \sigma }\sigma -g_{\Sigma \omega }\gamma _\mu
\omega ^\mu \right) \psi _\Sigma  \nonumber \\
& - & \bar \Sigma _{ij} \left( \frac{g_{\Sigma \rho }}{2}\gamma _\mu
\Theta ^\mu _{jk}+\frac{e}{2}\gamma _\mu A^\mu \left( \tau _3\right) _{jk}
\right) \Sigma _{ki}, \label{eq1}
\end{eqnarray}
where $\Sigma _{ij}$ and $\Theta _{ij}^\mu $ are the traceless $2\times 2$
matrices
\begin{equation}
\Sigma _{ij} =
\left( \begin{array}{cc}
\psi _{\Sigma ^0} & \sqrt{2}\psi _{\Sigma ^+} \\
\sqrt{2}\psi _{\Sigma ^-} & -\psi _{\Sigma ^0}
\end{array} \right) ,
\end{equation}
and
\begin{equation}
\Theta ^\mu _{ij} =
\left( \begin{array}{cc}
\rho ^\mu _0 & \sqrt{2} \rho ^\mu _+ \\
\sqrt{2} \rho ^\mu _- & -\rho ^\mu _0
\end{array} \right) .
\end{equation}
The sum on $\Sigma $ in the first line of Eq.~(\ref{eq1}) is over the charge
states $\Sigma ^-$, $\Sigma ^0$, and $\Sigma ^+$.
The Euler-Lagrange equations then yield the Dirac equations for the $\Sigma $
hyperons:
\begin{equation}
\left( i\gamma _\mu \partial ^\mu -M^\ast _\Sigma -g_{\Sigma \omega }
\gamma _\mu \omega ^\mu -I_{3\Sigma }g_{\Sigma \rho }\gamma _\mu
\rho _0^\mu -I_{3\Sigma }e\gamma _\mu A^\mu \right) \psi _\Sigma =0,
\label{eq4}
\end{equation}
\begin{equation}
M^\ast _\Sigma =M_\Sigma -g_{\Sigma \sigma }\sigma , \label{eq5}
\end{equation}
where $I_{3\Sigma }$ denotes the third isospin component; i.e.
$I_{3\Sigma }=-1,0,+1$ for $\Sigma ^-$, $\Sigma ^0$, and $\Sigma ^+$,
respectively. This means, that Eq.~(\ref{eq4}) already considers the fact,
that within a RHA only the charge neutral component of the $\rho $-meson
field, $\rho _0^\mu $, yields a nonzero ground-state expectation value.

The $\Lambda $ hyperon has isospin and charge zero and therefore cannot
couple to the $\rho $-meson and electromagnetic fields. Hence, under the
consideration of the universal hyperon coupling and the replacement of
$M_\Sigma $ by $M_\Lambda $ in Eq.~(\ref{eq5}), the $\Lambda $ Dirac equation
equals the $\Sigma ^0$ Dirac equation.

For calculating the hypernuclear spectra we made use of a technique similar to
the so-called expectation-value method, which was successfully used within
nonrelativistic and relativistic nuclear physics to incorporate shell effects
into semiclassical densities and energies (see Refs.
[12--14]): The Dirac-Hartree equations for the hyperons are solved only once
with the meson fields of the corresponding nucleonic system, which are
self-consistently determined within a relativistic
Thomas-Fermi (RTF) approximation, as an input.

To check the validity of this approximation we recalculated the $\Lambda $
single-particle spectra for the hypernuclei $^{40}_\Lambda $Ca and
$^{208}_\Lambda $Pb with the parameters of Ref. \cite{three} (parameter set
I in Table I). The results for various $\Lambda $ levels are displayed
in Table II where our results are denoted by $H^\ast $. Compared with the
fully self-consistent RHA results the $H^\ast $ approximation systematically
underestimates the $\Lambda $ bindings which may be attributed to a surface
energy that is somewhat too large within the RTF approach \cite{fifteen}.
But as expected, the agreement is better for the larger mass number $A$
and the deeper lying levels because in both cases the Thomas-Fermi assumption
of locally constant fields is more valid. In conclusion, our results show a
rather good agreement with those of Ref. \cite{three}, which gives confidence
in the described scheme.

In the next step we calculated several $\Lambda $ and $\Sigma $ hypernuclei
using a set of coupling constants from Ref. \cite{four}, which considers the
constraints (1)--(3) mentioned above and in addition has been successfully
used in the description of nuclear matter properties (parameter set II in
Table I). In Fig.1 we show the contributions of the meson and electromagnetic
fields to the hyperon self-energy for the nuclei $^{28}$Si, $^{40}$Ca,
$^{90}$Zr, and $^{208}$Pb. The nonrelativistic reduction of
the hyperon potential, (for $\Lambda $ and
$\Sigma ^0$ entirely, for $\Sigma ^\pm $ mainly) given by the difference
$g_{H\omega }\omega ^0-g_{H\sigma }\sigma ,H=\Lambda ,\Sigma $, is also
displayed. It is worth noting, that the small potential depths of
$\sim $~30~MeV go along with a relatively smooth radial dependence (compared
with nucleonic potentials), thereby additionally supporting the feasibility
of the Thomas-Fermi meson fields. A similar behaviour was found for
$^{16}_\Lambda$O within the RHA calculations of Ref. \cite{three}.
In Figs.2 and 3 we show the single-particle spectra of protons, neutrons,
$\Lambda $, $\Sigma ^0$, and $\Sigma ^+$ hyperons for the nuclei
$^{40}$Ca and $^{208}$Pb. Because of the smaller couplings the hyperon levels
are considerably less bound than the corresponding nucleon levels. Looking at
the $\Lambda $ and $\Sigma ^0$ single-particle energies, the larger
$M_\Sigma $ yields a smaller repulsive effect of the kinetic energy resulting
in systematically stronger bindings for the $\Sigma ^0$.

Dealing with hypernuclear states and their structure, one of the most
interesting questions concerns the spin-orbit potential for the hyperons
\cite{seven}. It is one of the great advantages of a relativistic treatment
that the spin-orbit interaction is automatically included in the
single-particle Dirac equation, and can be identified by means of a
Foldy-Wouthuysen reduction. For example, looking at the charge neutral
$\Lambda $ and $\Sigma ^0$ hyperons, both of which are not coupled to
the $\rho $-meson field, the ratio of the spin-orbit splitting
(Thomas terms) is
\begin{equation}
\frac{V_{s.o.}^{\Sigma ^0}}{V_{s.o.}^{\Lambda }}=
\frac{M^2_\Lambda }{M^2_\Sigma }=0.88\enspace . \label{eq6}
\end{equation}
This ratio is very well reproduced in the corresponding spectra of Figs. 2
and 3. Concerning the ratios of the spin-orbit splitting of the proton and
neutron to the $\Sigma ^+$ and $\Sigma ^0$ hyperon, respectively, a simple
expression similar to the one of Eq.~(\ref{eq6}) cannot be given because
of the different couplings. However, in the corresponding spectra of
Figs. 2 and 3 we found a value of about $\sim $~0.34.

In Table III we compare various $\Sigma ^-$, $\Sigma ^0$, and $\Sigma ^+$
levels for the $^{28}_\Sigma $Si, $^{40}_\Sigma $Ca, $^{90}_\Sigma $Zr,
and $^{208}_\Sigma $Pb hypernuclei. Of course, the Coulomb force plays an
important role: comparing $\Sigma ^-$ with $\Sigma ^0$ levels, the atomic
states disappear, whereas for the $\Sigma ^+$ states the baryonic potential
has to overcome the Coulomb repulsion with the effect that only the deep
lying states survive. For symmetric ($N=Z$) hypernuclei, where the
$\rho $-meson field is weak (it is nonzero because the proton and neutron
density distributions differ due to the Coulomb interaction),
we found the
Coulomb shifts between $\Sigma ^-$ and $\Sigma ^0$, or $\Sigma ^0$ and
$\Sigma ^+$ states almost identical to the corresponding neutron-proton
shifts in ``normal'' symmetric nuclei. The situation is somewhat different
for hypernuclei with a neutron-excess because then the effect of the
$\rho $-meson field is not negligible anymore: For the $\Sigma ^-$
($\Sigma ^+$) the $\rho $-meson adds (subtracts) from the isoscalar part
of the timelike repulsive vector field. The
$\Sigma ^0$ does not couple to the $\rho $ field at all. To get an idea of the
impact of the $\rho $-meson field we recalculated the asymmetric hypernuclei
$^{90}_\Sigma $Zr and $^{208}_\Sigma $Pb with the $\rho $-$\Sigma $ coupling
switched off. For $^{90}_\Sigma $Zr we found the $\Sigma ^-(\Sigma ^+)$ states
stronger (weaker) bound by about 2.1--3.5 MeV; the corresponding range for
$^{208}_\Sigma $Pb is 4.4--6.9 MeV. The fact that such relatively large ranges
occur can be understood in terms of the rms-radii: The lower bounds are for
weakly bound states with large rms-radii (e.g. the $1g_{9/2}\ \Sigma ^-$ state
with $r_{rms}=5.40$~fm for $^{90}_\Sigma $Zr and the $3p_{1/2}\ \Sigma ^-$
state with $r_{rms}=6.97$~fm for $^{208}_\Sigma $Pb), while the upper bounds
correspond to deep lying states with small rms-radii (e.g. the $1s_{1/2}\
\Sigma ^-$ states with $r_{rms}=3.09$~fm for $^{90}_\Sigma $Zr and $r_{rms}=
3.68$~fm for $^{208}_\Sigma $Pb; the values for the rms-radii are from the
calculations with the $\rho $ field switched on).
As one can see, the influence of the $\rho $-meson field, whose
range is determined by its mass $m_\rho $, weakens with increasing radial
distances.
(For neutrons and protons the situation is similar to the $\Sigma ^--\Sigma ^+$
pair but the effective $\rho $-nucleon coupling is weaker.)

Hence, the quantum hadrodynamical treatment of hypernuclei offers, by the
possible inclusion of the $\rho $-meson field, a natural way to incorporate
an isospin dependence into the $\Sigma $ potential (i.e. a Lane potential),
which was pointed out by Dover in Ref. \cite{seven} to be one of the most
important questions of hypernuclear physics.

Finally, we show in Figs.4--6 the single-particle energies of the $\Sigma ^-$,
$\Sigma ^0$, and $\Sigma ^+$ hyperons, respectively, versus
$A^{-2/3}$, with, $A$
the mass
number of the
nuclei.
For the $\Sigma ^-$ (Fig.4) the attractive Coulomb
potential alone, irrespective of the strength of the short-range nuclear
potential, is enough to bind $\Sigma ^-$ states. Some of these states
(the most bound) are such that the rms-radii of the $\Sigma ^-$ wavefunctions
are essentially inside the nucleus: Looking at $^{40}_{\Sigma ^-}$Ca
we found for the rms-radii of the plotted $s_{\Sigma ^-}$, $p_{\Sigma ^-}$,
$d_{\Sigma ^-}$, and $f_{\Sigma ^-}$ states the values of $r_{rms}$~=~2.65,
3.31, 3.89, and 4.76~fm, respectively. Therefore, compared with the
experimantal
charge rms-radius
of $r_c \sim 3.48$~fm for $^{40}$Ca, the $d_{\Sigma ^-}$ and
$f_{\Sigma ^-}$ states should be called {\it atomic} ones, while for
the other levels an identification as {\it hypernuclear} states seems to be
more appropriate.

For the charge neutral $\Sigma ^0$ (Fig.5) there is no Coulomb attraction
and therefore the number of bound states decreases. The value of -28~MeV
for $A^{-2/3}$~=~0.0 represents the binding energy of the lowest $\Sigma ^0$
level in nuclear matter, which is under the assumption of an universal hyperon
coupling the same as for the $\Lambda $ particle \cite{one,four}.
As expected, the pattern of states shows the standard behaviour as for
$\Lambda $ hypernuclei \cite{three}.

Turning finally to the discussion of the $A$-dependence of the $\Sigma ^+$
levels (Fig.6) the situation becomes more complicated. Of course, now the
Coulomb force is repulsive and the number of bound states further
decreases compared with $\Sigma ^-$ and $\Sigma ^0$ hypernuclei.
But to get a full understanding of Fig.6, it seems necessary to consider
the various contributions to the nuclear potential in view of their range,
which is determined by the corresponding meson mass, respectively. For
example, the fact that the binding of the $d_{\Sigma ^+}$ state increases
going from $A$~=~90 to $A$~=~208, while there is an opposite effect for the
$s_{\Sigma ^+}$ and $p_{\Sigma ^+}$ states, may be attributed to the larger
$d_{\Sigma ^+}$ rms-radius (e.g. 5.63~fm instead of 4.55~fm and 5.17~fm
for the $s_{\Sigma ^+}$ and $p_{\Sigma ^+}$ levels in $^{208}_{\Sigma ^+}$Pb,
respectively). The $d_{\Sigma ^+}$ wavefunction is located at large radial
distance, where the impact of the attractive $\sigma $-meson
($m_\sigma =500$~MeV) increases locally compared with the repulsive
$\omega $-meson ($m_\omega =783$~MeV) due to their different ranges.
Seemingly, the various states show a behaviour similar
to the one found within nonrelativistic Hartree-Fock Skyrme calculations
for protons in ``normal'' nuclei, where the binding of the $d$ levels strongly
increases, whereas the $s$ states nearly stay constant when going from
$A$~=90 to $A$~=208 \cite{sixteen}.

In the present calculations the broadening of the $\Sigma $ hyperon states due
to their decay to the $\Lambda $ was neglected. In principle, the model can be
extended to include the decay by introducing appropriate
$\Sigma \Lambda $-vertices. An important aspect of such an extended approach
would be the possibility of investigating
the decay of strange particles in the
nuclear medium. In order to estimate the effects due to the conversion
$\Sigma N\rightarrow \Lambda N$ the results of nonrelativistic potential
models \cite{seventeen} may be taken as a guideline. In such approaches the
decay is described schematically by an absorption potential for hyperons.
The imaginary part of the self-energy effectively lowers the
binding energy which can be understood in terms of the pole structure
of the baryon propagator. Qualitatively a similar effect has to be expected
also in a covariant description including the decay of the $\Sigma $
hyperons. Thus the present results are likely to give lower bounds for the
binding properties of the $\Sigma $ particles in hypernuclei.

In conclusion, we performed relativistic mean-field calculations of $\Lambda $
and $\Sigma $ hypernuclei using an interaction that considers neutron-star
masses, the $\Lambda $ binding in saturated nuclear matter, and experimental
$\Lambda $ single-particle levels. Concerning the $\Sigma $ couplings we
assumed an universal hyperon coupling; i.e. all hyperons in the lowest octet
couple to the meson fields as the $\Lambda $.
We employed the so-called
expectation-value method whose reliability was found to be sufficient compared
with fully self-consistent RHA calculations. Analyzing the hypernuclear
spectra, we devoted special attention to the spin-orbit interaction for
hyperons and, in the case of $\Sigma $ hypernuclei, to the isospin
dependence of the interaction. These two features are of particular interest
in the current discussion concerning hyperon potentials in nuclei and are
naturally incorporated into the relativistic quantum hadrodynamical model
we used. In addition we found by comparison with corresponding nonrelativistic
results that for strongly bound states the impact of the $\Sigma N\rightarrow
\Lambda N$ decay width on the bindings seems to be negligible.

In the future it would be very valuable from the point of view of dense matter
properties, and especially the structure of neutron-stars, to have the
assumption of an universal hyperon coupling confirmed by detailed precision
experiments on $\Sigma $ hypernuclei, and we hope that these calculations
may possibly be of assistence as well as a stimulus to such experiments
and the development of the necessary facilities.

\hfill \smallskip \\
{\large {\bf {Acknowledgments}}} \hfill \smallskip \\
One of us (D.~V.-E.) would like to thank the
Ludwig-Maximilians-Universit\"at M\"unchen for financial support.
The authors wish to thank Prof. W.~Stocker for helpful discussions
on the role of the surface energy.

\newpage
{\Large{\bf Table captions}}\\
\begin{description}
\item[table I:]
Parameters of the two forces considered in the text. In both cases the nucleon
mass $M=938$ MeV. For saturated nuclear matter the set II \cite{four} yields
energy per particle
$E/A=-16.3$ MeV, density $\rho _0=0.153$ fm$^{-3}$, incompressibility
$K=300$ MeV, effective mass $M^{\ast}/M=0.7$ and symmetry energy
coefficient $a_{sym}=32.5$ MeV. \\
$C_i^2=g_i^2\left( M/m_i \right) ^2,x_i=g_{Hi}/g_i; i=\sigma , \omega ,
\rho ; H=\Lambda , \Sigma .$
\item[table II:]
Comparison of our H$^{\ast }$ results for various $\Lambda $ levels in
$^{40}_\Lambda $Ca and $^{208}_\Lambda $Pb with the fully self-consistent
relativistic Hartree results from Ref. \cite{three}. All quantities are in MeV.
\item[table III:]
Various $\Sigma ^{-}$, $\Sigma ^{0}$, and $\Sigma ^{+}$ levels
in $^{28}_\Sigma $Si, $^{40}_\Sigma $Ca, $^{90}_\Sigma $Zr, and
$^{208}_\Sigma $Pb. All quantities are in MeV.
\end{description}
\newpage
\begin{center}
{\Large{\bf Table I}}
\end{center}
\vspace{0.5cm}
\begin{center}
\begin{tabular}{lcc}
\hline
& & \\
& I \cite{three} & II \cite{four} \\
& & \\
\hline
& & \\
$m_\sigma $~(MeV) & 499.31 & 500 \\
$m_\omega $~(MeV) & 780 & 783 \\
$m_\rho $~(MeV) & 763 & 770 \\
$M_\Lambda $~(MeV) & 1116.08 & 1115 \\
$M_\Sigma $~(MeV) & -- & 1190 \\
$C_\sigma ^2$ & 348.26 & 266.40 \\
$C_\omega ^2$ & 229.29 & 161.53 \\
$C_\rho ^2$ & 148.92 & 99.67 \\
$b\times 10^3$ & 2.2847 & 2.947 \\
$c\times 10^3$ & -2.9151 & -1.070 \\
$x_\sigma $ & 0.464 & 0.600 \\
$x_\omega $ & 0.481 & 0.653 \\
$x_\rho $ & -- & 0.600 \\
& & \\
\hline
\end{tabular}
\end{center}
\newpage
\begin{center}
{\Large{\bf Table II}}
\end{center}
\vspace{0.5cm}
\begin{center}
\begin{tabular}{crrrr}
\hline
& & & & \\
&
\multicolumn{2}{c}{\bf{$^{40}_\Lambda $Ca}} &
\multicolumn{2}{c}{\bf{$^{208}_\Lambda $Pb}} \\
\multicolumn{1}{c}{Level} &
\multicolumn{1}{c}{RHA \cite{three}} &
\multicolumn{1}{c}{H$^{\ast }$} &
\multicolumn{1}{c}{RHA \cite{three}} &
\multicolumn{1}{c}{H$^{\ast }$} \\
& & & & \\
\hline
& & & & \\
$1d_{3/2}$ & -2.63 & -1.17 & -15.78 & -15.03 \\
$1d_{5/2}$ & -3.76 & -2.08 & -16.12 & -15.32 \\
$1p_{1/2}$ & -10.93 & -8.75 & -20.38 & -19.42 \\
$1p_{3/2}$ & -11.61 & -9.38 & -20.51 & -19.54 \\
$1s_{1/2}$ & -19.43 & -16.90 & -24.19 & -23.23 \\
& & & & \\
\hline
\end{tabular}
\end{center}
\newpage
\begin{center}
{\Large{\bf Table III}}
\end{center}
\vspace{0.5cm}
\begin{center}
\begin{tabular}{crrrrrr}
\hline
& & & & & & \\
&
\multicolumn{3}{c}{\bf{$^{28}_\Sigma $Si}} &
\multicolumn{3}{c}{\bf{$^{40}_\Sigma $Ca}} \\
\multicolumn{1}{c}{Level} &
\multicolumn{1}{c}{$\Sigma ^{-}$} &
\multicolumn{1}{c}{$\Sigma ^{0}$} &
\multicolumn{1}{c}{$\Sigma ^{+}$} &
\multicolumn{1}{c}{$\Sigma ^{-}$} &
\multicolumn{1}{c}{$\Sigma ^{0}$} &
\multicolumn{1}{c}{$\Sigma ^{+}$} \\
& & & & & & \\
\hline
& & & & & & \\
$1f_{5/2}$ & -- & -- & -- & -2.31 & -- & -- \\
$1f_{7/2}$ & -- & -- & -- & -3.43 & -- & -- \\
$2s_{1/2}$ & -5.19 & -0.43 & -- & -9.73 & -2.57 & -- \\
$1d_{3/2}$ & -5.10 & -- & -- & -10.29 & -3.00 & -- \\
$1d_{5/2}$ & -6.19 & -0.87 & -- & -11.31 & -3.99 & -- \\
$1p_{1/2}$ & -13.97 & -7.96 & -2.07 & -18.79 & -10.82 & -3.00 \\
$1p_{3/2}$ & -14.71 & -8.69 & -2.79 & -19.37 & -11.42 & -3.60 \\
$1s_{1/2}$ & -23.22 & -16.63 & -10.11 & -27.10 & -18.54 & -10.07 \\
& & & & & & \\
&
\multicolumn{3}{c}{\bf{$^{90}_\Sigma $Zr}} &
\multicolumn{3}{c}{\bf{$^{208}_\Sigma $Pb}} \\
\hline
& & & & & & \\
$1f_{5/2}$ & -13.35 & -4.03 & -- & -26.47 & -12.40 & -- \\
$1f_{7/2}$ & -14.30 & -5.00 & -- & -26.87 & -12.87 & -- \\
$2s_{1/2}$ & -19.35 & -9.28 & -- & -30.79 & -15.76 & -0.73 \\
$1d_{3/2}$ & -20.29 & -10.52 & -0.87 & -31.57 & -16.99 & -2.70 \\
$1d_{5/2}$ & -20.88 & -11.15 & -1.53 & -31.76 & -17.24 & -3.02 \\
$1p_{1/2}$ & -26.90 & -16.61 & -6.47 & -36.25 & -21.06 & -6.22 \\
$1p_{3/2}$ & -27.16 & -16.91 & -6.80 & -36.31 & -21.15 & -6.36 \\
$1s_{1/2}$ & -32.94 & -22.05 & -11.31 & -40.44 & -24.48 & -8.94 \\
& & & & & & \\
\hline
\end{tabular}
\end{center}
\newpage
{\Large{\bf Figure captions}}\\
\begin{description}
\item[figure 1:]
Hyperon self-energy contributions $g_{H\sigma }\sigma $ (dotted lines),
$g_{H\omega }\omega ^0$ (dashed lines), $g_{\Sigma \rho }\rho _0^0$
(long-dashed lines), and $eA^0$ (dot-dashed lines) for the nuclei
$^{28}$Si, $^{40}$Ca, $^{90}$Zr, and $^{208}$Pb. The nonrelativistic
reduction of the hyperon potential
is (for $\Lambda $ and $\Sigma ^0$ entirely, for $\Sigma ^\pm$ mainly)
given by the difference $g_{H\omega }\omega ^0-g_{H\sigma }\sigma $,
which is represented by the solid lines, respectively. $H=\Lambda ,
\Sigma $ in the hyperon coupling constants.
\item[figure 2:]
The calculated proton, $\Sigma ^{+}$, neutron, $\Lambda $, and
$\Sigma ^{0}$ single-particle spectrum for $^{40}$Ca (parameter set II
of Table I).
\item[figure 3:]
Same as Fig.2 for $^{208}$Pb.
\item[figure 4:]
Single-particle energies versus $A^{-2/3}$ for $\Sigma ^{-}$.
For each angular momentum
the lowest lying state is plotted, respectively. The dashed lines are added
to guide the eye.
\item[figure 5:]
Same as Fig.4 for $\Sigma ^0$.
The value of
-28 MeV for $A^{-2/3}=0.0$ represents the binding energy of the lowest
$\Sigma ^{0}$ level in saturated nuclear matter under the assumption of an
universal hyperon coupling \cite{one,four}.
\item[figure 6:]
Same as Fig.4 for $\Sigma ^+$.
\end{description}
\newpage
\pagestyle{empty}
\begin{figure}
\begin{center}
\unitlength0.45cm
\begin{picture}(23.5,40)
\thicklines
\put(0.0,0.0){\framebox(23.5,40)}
\put(11.0,38.5){\bf{$^{40}$Ca}}
\put(3.8,37.0){$p$} \put(7.7,37.0){$\Sigma ^{+}$}
\put(13.3,37.0){$n$} \put(17.3,37.0){$\Lambda $}
\put(21.2,37.0){$\Sigma ^{0}$}
\put(9.5,0.0){\line(0,1){36.5}}
\put(0.0,36.5){\line(1,0){23.5}}
\multiput(0.0,36.0)(0.0,-5.0){8}{\line(1,0){0.3}}
\multiput(0.0,36.0)(0.0,-1.0){36}{\line(1,0){0.15}}
\multiput(9.5,36.0)(0.0,-5.0){8}{\line(1,0){0.3}}
\multiput(9.5,36.0)(0.0,-1.0){36}{\line(1,0){0.15}}
\put(-4.0,35.75){MeV}
\put(-1.7,35.75){0.0}
\put(-2.0,30.75){-5.0}
\put(-2.5,25.75){-10.0}
\put(-2.5,20.75){-15.0}
\put(-2.5,15.75){-20.0}
\put(-2.5,10.75){-25.0}
\put(-2.5,5.75){-30.0}
\put(-2.5,0.75){-35.0}
\put(3.0,26.42){\line(1,0){2.0}}
\put(1.0,26.17){$2s_{1/2}$}
\put(1.0,10.03){$1p_{1/2}$}
\put(3.0,10.28){\line(1,0){2.0}}
\put(3.0,8.31){\line(1,0){2.0}}
\put(1.0,8.06){$1p_{3/2}$}
\put(3.0,25.02){\line(1,0){2.0}}
\put(1.0,24.77){$1d_{3/2}$}
\put(3.0,21.51){\line(1,0){2.0}}
\put(1.0,21.26){$1d_{5/2}$}
\put(3.0,34.72){\line(1,0){2.0}}
\put(1.0,34.47){$1f_{7/2}$}
\put(8.0,25.93){\line(1,0){1.0}}
\put(6.0,25.68){$1s_{1/2}$}
\put(8.0,33.00){\line(1,0){1.0}}
\put(6.0,32.95){$1p_{1/2}$}
\put(8.0,32.40){\line(1,0){1.0}}
\put(6.0,32.10){$1p_{3/2}$}
\put(12.5,18.75){\line(1,0){2.0}}
\put(10.5,18.50){$2s_{1/2}$}
\put(12.5,2.22){\line(1,0){2.0}}
\put(10.5,1.97){$1p_{1/2}$}
\put(10.5,33.10){$2p_{1/2}$}
\put(12.5,33.25){\line(1,0){2.0}}
\put(12.5,0.28){\line(1,0){2.0}}
\put(10.5,0.28){$1p_{3/2}$}
\put(12.5,32.21){\line(1,0){2.0}}
\put(10.5,32.36){$2p_{3/2}$}
\put(12.5,17.34){\line(1,0){2.0}}
\put(10.5,17.09){$1d_{3/2}$}
\put(12.5,13.83){\line(1,0){2.0}}
\put(10.5,13.58){$1d_{5/2}$}
\put(12.5,32.07){\line(1,0){2.0}}
\put(10.5,31.52){$1f_{5/2}$}
\put(12.5,27.45){\line(1,0){2.0}}
\put(10.5,27.20){$1f_{7/2}$}
\put(17.5,17.85){\line(1,0){1.0}}
\put(15.5,17.60){$1s_{1/2}$}
\put(17.5,34.23){\line(1,0){1.0}}
\put(15.5,34.28){$2s_{1/2}$}
\put(17.5,25.94){\line(1,0){1.0}}
\put(15.5,25.69){$1p_{1/2}$}
\put(17.5,25.25){\line(1,0){1.0}}
\put(15.5,25.00){$1p_{3/2}$}
\put(17.5,33.98){\line(1,0){1.0}}
\put(15.5,33.58){$1d_{3/2}$}
\put(17.5,32.89){\line(1,0){1.0}}
\put(15.5,32.64){$1d_{5/2}$}
\put(20.5,17.46){\line(1,0){2.0}}
\put(20.5,33.43){\line(1,0){2.0}}
\put(20.5,25.18){\line(1,0){2.0}}
\put(20.5,24.58){\line(1,0){2.0}}
\put(20.5,33.00){\line(1,0){2.0}}
\put(20.5,32.01){\line(1,0){2.0}}
\end{picture}
\end{center}
\end{figure}
\newpage
\begin{figure}
\begin{center}
\unitlength0.45cm
\begin{picture}(23.5,40)
\thicklines
\put(0.0,0.0){\framebox(23.5,40)}
\put(11.0,38.5){\bf{$^{208}$Pb}}
\put(3.8,37.0){$p$} \put(7.7,37.0){$\Sigma ^{+}$}
\put(13.3,37.0){$n$} \put(17.3,37.0){$\Lambda $}
\put(21.2,37.0){$\Sigma ^{0}$}
\put(9.5,0.0){\line(0,1){36.5}}
\put(0.0,36.5){\line(1,0){23.5}}
\multiput(0.0,36.0)(0.0,-5.0){8}{\line(1,0){0.3}}
\multiput(0.0,36.0)(0.0,-1.0){36}{\line(1,0){0.15}}
\multiput(9.5,36.0)(0.0,-5.0){8}{\line(1,0){0.3}}
\multiput(9.5,36.0)(0.0,-1.0){36}{\line(1,0){0.15}}
\put(-4.0,35.75){MeV}
\put(-1.7,35.75){0.0}
\put(-2.0,30.75){-5.0}
\put(-2.5,25.75){-10.0}
\put(-2.5,20.75){-15.0}
\put(-2.5,15.75){-20.0}
\put(-2.5,10.75){-25.0}
\put(-2.5,5.75){-30.0}
\put(-2.5,0.75){-35.0}
\put(3.0,8.62){\line(1,0){2.0}}
\put(1.0,8.37){$2s_{1/2}$}
\put(1.0,28.03){$3s_{1/2}$}
\put(3.0,28.28){\line(1,0){2.0}}
\put(3.0,17.47){\line(1,0){2.0}}
\put(1.0,17.45){$2p_{1/2}$}
\put(3.0,17.00){\line(1,0){2.0}}
\put(1.0,16.75){$2p_{3/2}$}
\put(3.0,5.67){\line(1,0){2.0}}
\put(1.0,5.42){$1d_{3/2}$}
\put(3.0,4.88){\line(1,0){2.0}}
\put(3.0,25.78){\line(1,0){2.0}}
\put(3.0,13.14){\line(1,0){2.0}}
\put(3.0,11.72){\line(1,0){2.0}}
\put(3.0,34.79){\line(1,0){2.0}}
\put(3.0,21.44){\line(1,0){2.0}}
\put(3.0,19.25){\line(1,0){2.0}}
\put(3.0,30.35){\line(1,0){2.0}}
\put(3.0,27.33){\line(1,0){2.0}}
\put(3.0,35.81){\line(1,0){2.0}}
\put(3.0,26.66){\line(1,0){2.0}}
\put(1.0,4.63){$1d_{5/2}$}
\put(1.0,25.53){$2d_{5/2}$}
\put(1.0,12.89){$1f_{5/2}$}
\put(1.0,11.47){$1f_{7/2}$}
\put(1.0,34.54){$2f_{7/2}$}
\put(1.0,21.19){$1g_{7/2}$}
\put(1.0,19.00){$1g_{9/2}$}
\put(1.0,30.10){$1h_{9/2}$}
\put(0.75,27.29){$1h_{11/2}$}
\put(0.86,35.56){$1i_{13/2}$}
\put(1.0,26.41){$2d_{3/2}$}
\put(8.0,27.06){\line(1,0){1.0}}
\put(6.0,26.81){$1s_{1/2}$}
\put(8.0,35.27){\line(1,0){1.0}}
\put(6.0,35.02){$2s_{1/2}$}
\put(8.0,29.78){\line(1,0){1.0}}
\put(6.0,29.95){$1p_{1/2}$}
\put(8.0,29.64){\line(1,0){1.0}}
\put(6.0,29.30){$1p_{3/2}$}
\put(8.0,33.30){\line(1,0){1.0}}
\put(6.0,33.42){$1d_{3/2}$}
\put(8.0,32.98){\line(1,0){1.0}}
\put(6.0,32.65){$1d_{5/2}$}
\put(12.5,17.33){\line(1,0){2.0}}
\put(10.5,16.93){$3s_{1/2}$}
\put(12.5,35.65){\line(1,0){2.0}}
\put(10.5,35.86){$4s_{1/2}$}
\put(10.5,6.60){$2p_{1/2}$}
\put(12.5,6.64){\line(1,0){2.0}}
\put(12.5,27.40){\line(1,0){2.0}}
\put(10.5,27.60){$3p_{1/2}$}
\put(12.5,6.17){\line(1,0){2.0}}
\put(10.5,5.92){$2p_{3/2}$}
\put(12.5,26.97){\line(1,0){2.0}}
\put(10.5,26.87){$3p_{3/2}$}
\put(12.5,15.97){\line(1,0){2.0}}
\put(10.5,15.72){$2d_{3/2}$}
\put(12.5,35.28){\line(1,0){2.0}}
\put(12.5,2.85){\line(1,0){2.0}}
\put(12.5,25.40){\line(1,0){2.0}}
\put(12.5,1.52){\line(1,0){2.0}}
\put(12.5,24.11){\line(1,0){2.0}}
\put(12.5,11.35){\line(1,0){2.0}}
\put(12.5,34.43){\line(1,0){2.0}}
\put(12.5,9.25){\line(1,0){2.0}}
\put(12.5,32.93){\line(1,0){2.0}}
\put(12.5,20.41){\line(1,0){2.0}}
\put(12.5,17.46){\line(1,0){2.0}}
\put(12.5,26.00){\line(1,0){2.0}}
\put(12.5,34.72){\line(1,0){2.0}}
\put(12.5,29.80){\line(1,0){2.0}}
\put(12.5,15.08){\line(1,0){2.0}}
\put(10.5,2.60){$1f_{5/2}$}
\put(10.5,25.10){$2f_{5/2}$}
\put(10.5,35.23){$3d_{5/2}$}
\put(10.5,1.27){$1f_{7/2}$}
\put(10.5,23.86){$2f_{7/2}$}
\put(10.5,11.10){$1g_{7/2}$}
\put(10.5,33.66){$2g_{7/2}$}
\put(10.5,9.00){$1g_{9/2}$}
\put(10.5,14.83){$2d_{5/2}$}
\put(10.5,32.68){$2g_{9/2}$}
\put(10.20,17.62){$1h_{11/2}$}
\put(10.36,34.47){$1j_{15/2}$}
\put(10.36,25.95){$1i_{13/2}$}
\put(10.36,29.55){$1i_{11/2}$}
\put(10.5,20.16){$1h_{9/2}$}
\put(17.5,11.72){\line(1,0){1.0}}
\put(15.5,11.47){$1s_{1/2}$}
\put(17.5,20.92){\line(1,0){1.0}}
\put(15.5,20.67){$2s_{1/2}$}
\put(17.5,32.59){\line(1,0){1.0}}
\put(15.5,32.77){$3s_{1/2}$}
\put(17.5,15.34){\line(1,0){1.0}}
\put(15.5,15.44){$1p_{1/2}$}
\put(17.5,26.45){\line(1,0){1.0}}
\put(15.5,26.60){$2p_{1/2}$}
\put(17.5,15.23){\line(1,0){1.0}}
\put(15.5,14.73){$1p_{3/2}$}
\put(17.5,26.26){\line(1,0){1.0}}
\put(15.5,25.90){$2p_{3/2}$}
\put(17.5,19.64){\line(1,0){1.0}}
\put(15.5,19.64){$1d_{3/2}$}
\put(17.5,31.99){\line(1,0){1.0}}
\put(15.5,31.99){$2d_{3/2}$}
\put(17.5,19.34){\line(1,0){1.0}}
\put(15.5,18.95){$1d_{5/2}$}
\put(17.5,31.68){\line(1,0){1.0}}
\put(15.5,31.20){$2d_{5/2}$}
\put(17.5,24.49){\line(1,0){1.0}}
\put(15.5,24.40){$1f_{5/2}$}
\put(17.5,23.93){\line(1,0){1.0}}
\put(15.5,23.60){$1f_{7/2}$}
\put(17.5,29.74){\line(1,0){1.0}}
\put(15.5,29.49){$1g_{7/2}$}
\put(17.5,28.89){\line(1,0){1.0}}
\put(15.5,28.64){$1g_{9/2}$}
\put(17.5,35.25){\line(1,0){1.0}}
\put(15.5,35.00){$1h_{9/2}$}
\put(17.5,34.12){\line(1,0){1.0}}
\put(15.25,33.87){$1h_{11/2}$}
\put(20.5,11.52){\line(1,0){2.0}}
\put(20.5,20.24){\line(1,0){2.0}}
\put(20.5,31.50){\line(1,0){2.0}}
\put(20.5,14.94){\line(1,0){2.0}}
\put(20.5,25.50){\line(1,0){2.0}}
\put(20.5,14.85){\line(1,0){2.0}}
\put(20.5,25.34){\line(1,0){2.0}}
\put(20.5,19.01){\line(1,0){2.0}}
\put(20.5,30.85){\line(1,0){2.0}}
\put(20.5,18.76){\line(1,0){2.0}}
\put(20.5,30.57){\line(1,0){2.0}}
\put(20.5,23.60){\line(1,0){2.0}}
\put(19.5,35.78){$2f_{5/2}$}
\put(21.5,35.93){\line(1,0){1.0}}
\put(20.5,23.13){\line(1,0){2.0}}
\put(19.5,35.07){$2f_{7/2}$}
\put(21.5,35.62){\line(1,0){1.0}}
\put(20.5,28.61){\line(1,0){2.0}}
\put(20.5,27.88){\line(1,0){2.0}}
\put(20.5,33.90){\line(1,0){2.0}}
\put(20.5,32.92){\line(1,0){2.0}}
\end{picture}
\end{center}
\end{figure}
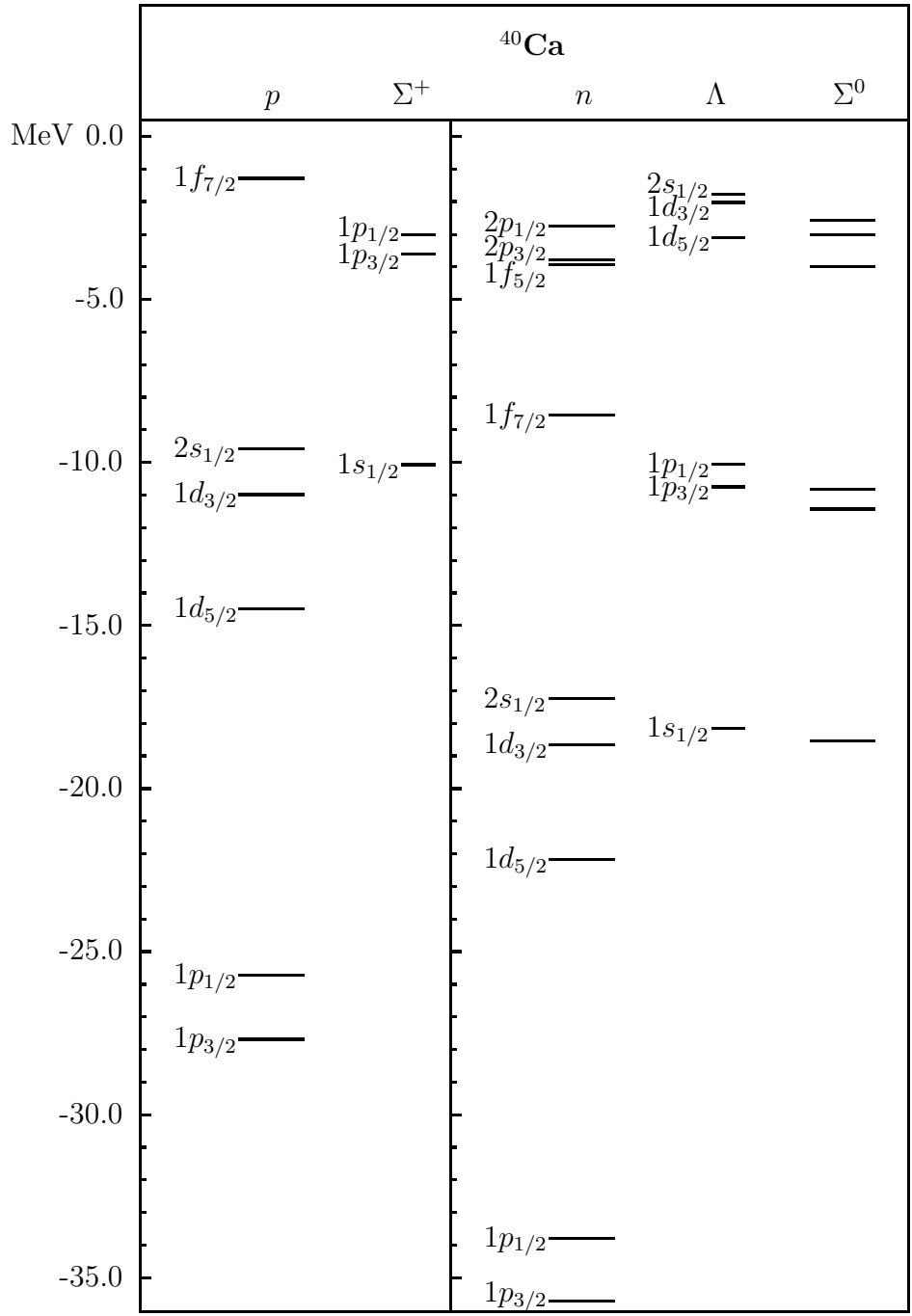
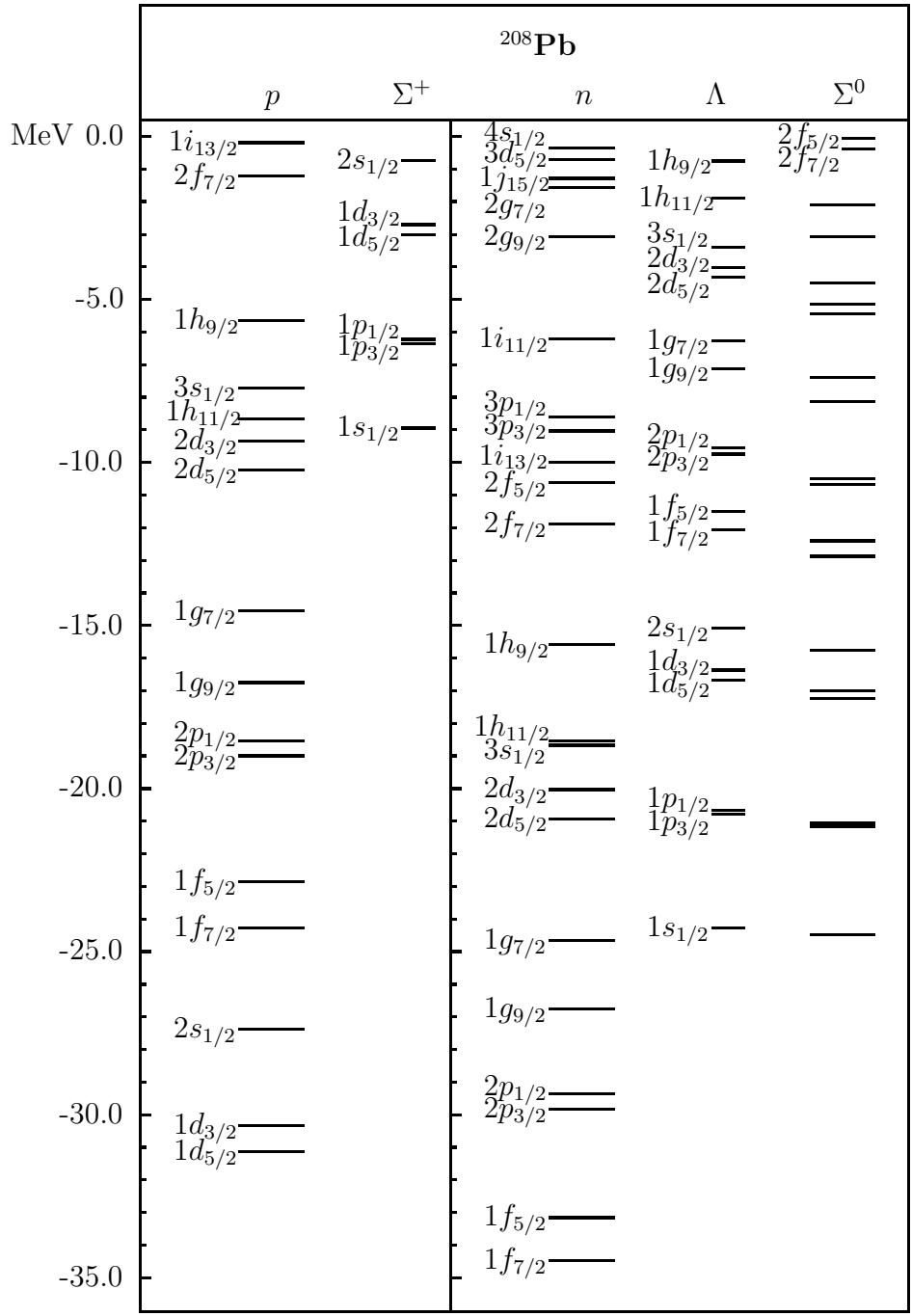
\end{document}